\documentclass[twocolumn,showpacs,nofootinbib,tightenlines,nobibnotes, aps, amssymb,amsmath,prb]{revtex4}
\usepackage{graphicx}
\usepackage{epsfig}

\newcommand{\la}{\langle}
\newcommand{\ra}{\rangle}

\newcommand{\nn}{\nonumber}

\newcommand{\be}{\begin{eqnarray}}
\newcommand{\ee}{\end{eqnarray}}

\begin{document}
\title{Persistent supersolid phase of hard-core bosons on the triangular lattice}
\author{Dariush Heidarian and Kedar Damle}
\affiliation{%
Department of Theoretical Physics,
Tata Institute of Fundamental Research,
Homi Bhabha Road, Colaba, Mumbai 400005, India
}%

\date{May 10, 2005}

\begin{abstract}
We study hard-core bosons with unfrustrated hopping ($t$) and
nearest neighbour repulsion ($U$) on the triangular
lattice. At half-filling, the system undergoes a zero temperature ($T$)
quantum phase transition from a superfluid
phase at small $U$ to a supersolid at $U_c \approx 4.45$ in units of $2t$. This
supersolid phase breaks the lattice translation symmetry in a
characteristic $\sqrt{3} \times \sqrt{3}$ pattern, and is
remarkably stable---indeed, a smooth extrapolation of our results indicates
that the supersolid
phase persists for arbitrarily large $U/t$.
\end{abstract}

\pacs{75.10.Jm 05.30.Jp 71.27.+a}
\vskip2pc

\maketitle

{\it{Introduction:}} 
The observation of strongly correlated Mott insulating states
and $T=0$ superfluid-insulator transitions of ultracold
bosonic atoms subjected to optical lattice potentials\cite{BEC1} has led
to a great deal of interest in strongly correlated lattice systems that
can be realized in such experiments.\cite{BEC2,BEC3}
The recent observation of a supersolid phase in Helium\cite{Chan0}
leads, in this context, to a natural question: Can the
lattice analog of this, namely a superfluid
phase that simultaneously breaks lattice translation symmetry,
be seen in atom-trap experiments?

One class of promising candidates are systems which are
superfluid when interactions are weak, but form
insulators with spatial symmetry breaking when
interactions are strong:
In terms of conventional Landau theory, a direct transition between
these two states is generically either first order, or pre-empted by an
intermediate supersolid phase with both order parameters nonzero; both types of behaviour are known to occur in specific lattice models.\cite{Hebert0,Troyer1,Frey0}
Moreover, as has been
shown recently by Senthil {\it et. al.},\cite{Senthil0}
conventional Landau
theory can fail in certain situations in which quantum mechanical Berry phase
effects produce a direct second-order phase transition, thereby ruling
out an intermediate supersolid phase. When such a transition
occurs,\cite{Melko0,Sandvik0} it is associated with quasi-particle
fractionalization and deconfinement, and this alternative to an
intermediate supersolid phase is thus interesting
in its own right.
 
Bosons on the triangular
lattice with on-site repulsion $V$,
repulsive nearest neighbour interaction $U$, and
unfrustrated hopping ($t$) provide a particularly interesting
example in this context since the structure of interactions is simple enough
that
it can be realized in
atom-trap
experiments.\cite{BEC2,BEC4} In the hard-core $V \rightarrow \infty$ limit
(which is also experimentally feasible\cite{BEC2,BEC4})
this maps to a system of $S=1/2$ spins ($S^z_i = n_i -1/2$ where $n_i$
is the boson number at site $i$) with
antiferromagnetic exchange $J^z = U$ between the $z$ components of
neighbouring spins,
ferromagnetic exchange $J_\perp = 2t$ between their $x$ and $y$ components,
and magnetic field in the $z$ direction equal to the chemical potential $\mu$.
It is this hard-core limit we consider in some detail below at zero
chemical potential.

\begin{figure}[t]
\includegraphics[width=.7\hsize]{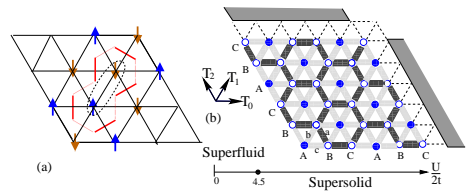}
\caption{a) A {\it flippable} pair of spins and the corresponding
hexagon-pair with its flippable dimer configuration. Also shown are
static spins that surround this flippable pair to form an elongated hexagonal
tile. Tiling the lattice with these tiles gives a candidate insulating state
at large $U/t$ ({\it not} observed numerically).
b) Actual $T=0$ phase diagram and nature of spatial symmetry breaking in the
supersolid phase. Darker bonds and sites indicate higher kinetic energy and
density respectively, and the state shown corresponds to
$\theta_K = \theta_n = 0$. Note that the lattice is drawn to
emphasize periodicity in directions
$T_0$ and $T_2$.}
\label{phases}
\end{figure}

Clearly,
the ground-state in the limit $U/t \rightarrow 0$ must be a
featureless superfluid. On the other hand,
the interaction energy $U$ dominates in the $U/t \rightarrow \infty$
limit and leads to frustration since
it is impossible to have all pairs of neighbouring spins pointing
anti-parallel to each other along the $z$ axis on the triangular lattice.
The ground state in this limit is thus expected to live entirely in the
highly degenerate
minimally frustrated subspace
of configurations with precisely one frustrated bond (parallel spins) per
triangle, and is selected by the dynamics associated with
the hopping term $t$. The minimally frustrated subspace can be
conveniently represented by noting that each state
in this subspace corresponds to a perfect dimer cover of the dual
hexagonal lattice (with every frustrated bond on the triangular lattice
mapping to a dimer placed on the dual link perpendicular to the bond
in question). In this language, the effective Hamiltonian in the
$U/t \rightarrow \infty$
limit is then a quantum dimer model with a ring-exchange term that
operates on each pair of adjacent hexagons of the dual lattice (see
Fig.~\ref{phases}).

Quantum dimer models with ring-exchange
on individual plaquettes of two dimensional bipartite lattices quite
generally have crystalline
ground states in which the spatial arrangement of dimers breaks
lattice symmetry.\cite{Sachdevreview,MoessnerPRB}
In our problem, a wavefunction that
gains kinetic energy from the double-hexagon ring exchange process
on a maximal set of {\it independently flippable} hexagon pairs
(see Fig.~\ref{phases}) provides a similar
candidate lattice symmetry breaking insulating state at large $U/t$
(see Fig.~\ref{phases}).
Note however that this analogy to simpler quantum dimer models misses
the important $U(1)$ symmetry associated with charge conservation.
Alternatively,
one can focus on this $U(1)$ symmetry at large $U/t$ by thinking in
terms of superfluid wavefunctions projected
into the minimally frustrated subspace: Clearly, superfluidity
can survive in such a projected state since the minimally
frustrated manifold admits considerable charge fluctuations, and
such wavefunctions also provide substantial kinetic energy gain.~\cite{Dhar}
The breaking of lattice translation symmetry that seems natural
by analogy to the simpler quantum dimer
models
then motivates a large-$U$ variational ground state obtained
by projecting a supersolid wavefunction.~\cite{Dhar}
This suggests that the `intermediate' supersolid phase of Landau theory may, in fact, persist to large $U$ in this case (another argument for a
supersolid was given in Ref.~\onlinecite{Murthy}).

While these considerations are not definitive, they do at least emphasize
that the behavior of this system at intermediate and
large $U$
presents very interesting possibilities, and a detailed numerical study is
one way to decide between them.
In the present work, we use Quantum Monte-Carlo (QMC) methods
to perform such a numerical study.
Our results for the different $T=0$ phases
are shown in Fig.~\ref{phases}. To summarize, we find that
the superfluid at small $U$ undergoes a transition to a
{\it supersolid} phase at $U_c \approx 4.45$ in units of $2t$.
This supersolid phase
breaks lattice translation symmetry in a characteristic
$ \sqrt{3} \times \sqrt{3}$ pattern shown in Fig.~\ref{phases}, and
appears
to be indeed stable for arbitrarily large values of $U/t$
(albeit with progressively smaller superfluid density).

{\it{Model and method:}}
Our Hamiltonian reads
\be
H &=& \sum_ {\la ij\ra}[U(n_{i}-1/2)(n_{j}-1/2) -t(b^{\dagger}_{i}b_{j} + b_{i} b^{\dagger}_{j})] \nn \\
&& + \sum_i [V(n_i -1/2)^2 -\mu n_i] \; ,
\label{eq:H_e}
\ee
where $\la ij\ra$ refer to nearest neighbour links of the two-dimensional
triangular lattice, $n_i$ is the particle number
at site $i$, and $b_i^{\dagger}$ is the boson creation operator
at site $i$. In this work, we take the limit $V \rightarrow \infty$
to enforce the hard-core constraint,
thereby mapping it to the $S=1/2$ spin model as mentioned earlier, set
$t$ to $1/2$, and take $\mu = 0$. 
We use the well-documented stochastic series expansion (SSE)
QMC method\cite{Syljuasen0,Sandvik2,Sandvik1} which
efficiently samples the high-temperature expansion
of the grand-canonical partition function.
(At large values of $U/t$, some modifications to the standard algorithm
are necessary, and these will be discussed separately\cite{unpublished}).

Most of our data is on $L \times L$ samples with periodic
boundary conditions and $L$ a multiple of six ranging from 12 to 48
at inverse temperatures $\beta$ ranging from $10$ to $30$.
Our
choice of boundary conditions and aspect ratio ensures that all the lattice
symmetries are preserved after imposing the boundary conditions
(see Fig.~\ref{phases}).
The nature of the $T\rightarrow 0$ phase and its low energy  spectrum
of excited states is conveniently characterized by the superfluid density
$\rho_s$, and the momentum ($\vec{q}$) and imaginary time ($\tau$)
dependent correlation functions $C_\rho(\vec{q},\tau)$,
$C_K^{\alpha \alpha^{'}}(\vec{q},\tau)$ 
of local particle density and kinetic energy respectively ($\alpha$
and $\alpha^{'}$ refer to the three possible bond
orientations $T_{0/1/2}$ shown in Fig.~\ref{phases}). We use standard SSE estimators\cite{Sandvik1} to calculate $\rho_s$,
$C_{\rho}(\vec{q},\tau=0)$,
$S_{\rho}(\vec{q},\omega_n=0) = \int_{0}^{\beta}d \tau C_{\rho}(\vec{q},\tau)$,
$C_{{\mathrm {K}}}^{\alpha \beta}(\vec{q},\tau=0)$ and 
$S_{{\mathrm {K}}}^{\alpha \beta}(\vec{q},\omega_n=0) = \int_{0}^{\beta}
d \tau C_{{\mathrm {K}}}(\vec{q},\tau)$.
These momentum space correlation functions are an unbiased probe of spatial
order in the system. By analyzing the $L$ and $\beta$ dependence of
the Bragg peaks at $\pm \vec{Q} = \pm (2\pi/3, 2\pi/3)$ seen in
the equal time and static correlation functions of density and kinetic
energy, we conclude that spatial order is established
at these wavevectors  when lattice
translation symmetry is broken in the supersolid phase (in the
convention used above, the components
of $\vec{Q}$ refer to projections in directions $T_0$ and
$T_2$ shown in Fig.~\ref{phases}.)  \begin{figure}[t]
\includegraphics[width=.8\hsize]{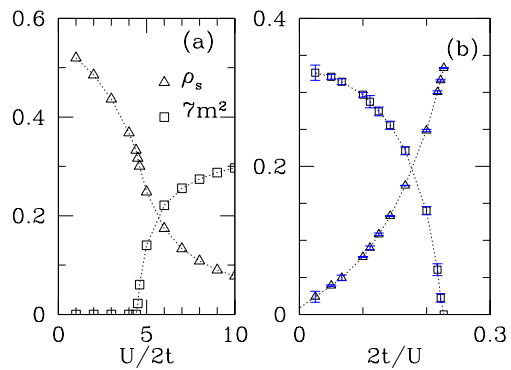}
\caption{Superfluid density $\rho_s$ and density wave order parameter
$m^2 \equiv S_{\rho}(\vec{Q},\omega_n=0)/\beta L^2$ extrapolated to
$T \rightarrow 0$ and $L \rightarrow \infty$.}
\label{rhosandnq}
\end{figure}
\begin{figure}[t]
\includegraphics[width=.8\hsize]{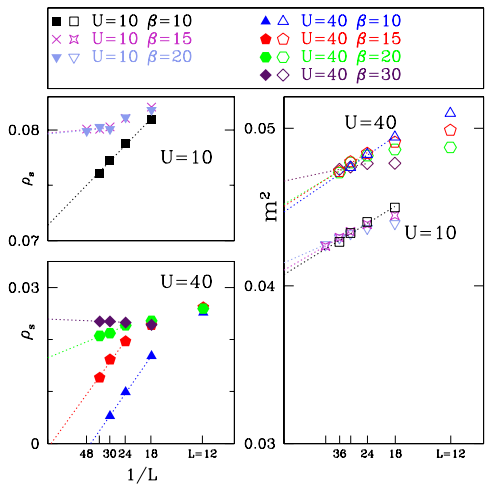}
\caption{Extrapolations implicit in Fig.~\ref{rhosandnq} shown here for two
values of $U$}
\label{extrapolations}
\end{figure}

We also measure two complex
order parameters sensitive to this spatial symmetry breaking to obtain
a better characterization of the supersolid state. These are defined as
\be
\psi_n & = & n_{{\mathrm A}} + n_{{\mathrm B}} \, e^{2\pi i/3} +
                   n_{{\mathrm C}} \, e^{4\pi i/3} \, , \nn \\
\psi_K & = & K_{{\mathrm a}} + K_{{\mathrm b}} \, e^{2\pi i/3} +
                   K_{{\mathrm c}} \, e^{4\pi i/3} \, .
\label{orderparameters}
\ee
Here the subscripts refer to the three-sublattice decomposition of
the triangular lattice into ${\mathrm A}$, ${\mathrm B}$, ${\mathrm C}$ type sites,
and ${\mathrm a}\equiv {\mathrm{BC}}$,
${\mathrm b} \equiv {\mathrm{CA}}$, ${\mathrm c} \equiv {\mathrm{AB}}$ type bonds respectively, while $n$ and $K$
are the densities and kinetic energies on the corresponding sites and
bonds. $\psi_K$ may be obtained
from the Fourier components of the kinetic energies in the three lattice
directions, $K_{\vec{Q}}^{(0/1/2)}$,
using the relation $\psi_K = e^{4\pi i/3} K_{\vec{Q}}^{(0)} +
e^{2\pi i/3} K_{\vec{Q}}^{(1)} + e^{4\pi i/3} K_{\vec{Q}}^{(2)}$, while
$\psi_n$ is precisely equal to $n_{\vec Q}$, the Fourier component of
the density at the ordering wavevector $\vec{Q}$. We expect both
order parameters to average to zero as long as our algorithm remains
ergodic---the probability distribution of their phases
$\theta_K$ and $\theta_n$ however provides useful
information regarding the nature of the supersolid phase.

{\it{Numerical results:}}
Our numerical results for the variation in the superfluid density $\rho_s$
as a function of $U$ are shown in Fig.~\ref{rhosandnq}. Each point shown in
Fig.~\ref{rhosandnq} represents an extrapolation of available data to the $T=0$
thermodynamic limit. The results summarized in
Fig.~\ref{rhosandnq} show no indication of any finite $U/t$ quantum
phase transition beyond which $\rho_s$ may become zero at zero temperature. Indeed,
a smooth extrapolation of our data suggests that superfluidity persists in the low
temperature limit at all finite values of $U$, albeit with an increasingly
small $T=0$ value of $\rho_s$. While this is surprising
from the perspective of putative Valence Bond Solid (VBS) ground states of
the corresponding quantum dimer model, further insight can be obtained
by performing a variational
calculation using projected superfluid wavefunctions; this work
will be reported separately.~\cite{Dhar}\begin{figure}[t]
\includegraphics[width=.7\hsize]{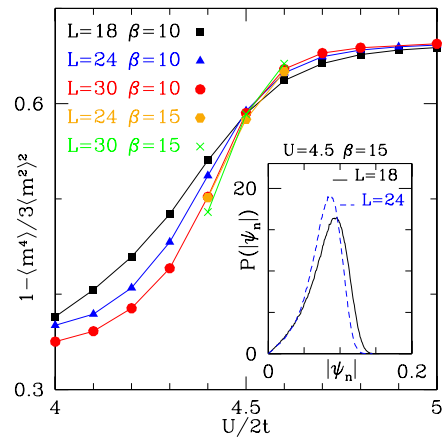}
\caption{Binder cumulant $g = 1 - \langle m^4 \rangle /3 \langle m^2
\rangle^2$ as a function of
$U$ in the transition region. From the location of the crossing point
we identify a $T=0$ phase transition to the supersolid phase at
$U \approx 4.45$. Inset: Histogram of $|\psi_n|$ has a single peak, indicating
a second-order transition}
\label{binder}
\end{figure}

\begin{figure}[t]
\includegraphics[width=.67\hsize]{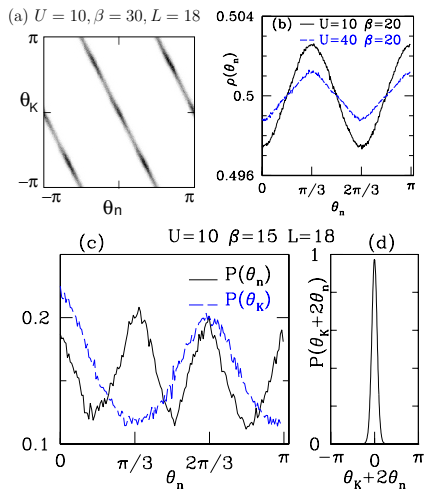}
\caption{Top panels: Greyscale plot of the joint probability distribution
of $\theta_n$ and $\theta_K$; $\theta_n$ dependence of $\rho$.
Bottom panels: Probability distribution of $\theta_n$
$\theta_K + 2\theta_n$, and $\theta_K$. The histogram of $\theta_K + 2\theta_n$
has additional peaks at $\pm 2\pi$ (not shown).}
\label{thetahistogram}
\end{figure}

Although superfluidity survives in the entire range of $U$ studied, the
state at small $U$  is not continuously connected to that at large $U$.
Indeed, we see clear evidence for a continuous $T=0$ quantum phase transition
at $U_c \approx 4.45$. This transition point is estimated using standard
criteria in terms of Binder cumulants as shown in Fig.~\ref{binder}.
(further details regarding the phase transition will be reported
separately\cite{unpublished}).
For $U>U_c$,
the system
spontaneously breaks lattice symmetry to produce a {\it supersolid} phase.
To understand the nature of the supersolid phase, it is useful to analyze the
joint probability distribution of the phases $\theta_K$ and $\theta_n$ defined
earlier. At $U=10$, we see from Fig.~\ref{thetahistogram} that $\theta_K$ is essentially
pinned to be equal to $-2\theta_n$ (modulo $2\pi$) at low temperature and large $L$, while
$\theta_n$ has a distribution which  peaks at $\theta_n^{p} = 2\pi p/6$
with $p$ an integer from $0$ to $5$. The picture that emerges (and gets
sharper at larger $L$ and $\beta$)
is thus of a state in which a relatively more mobile fluid of 
density $\rho_{{\mathrm{hx}}}$ living on
a hexagonal lattice backbone of the full lattice is responsible for the
superfluidity, while the centers of the hexagons have an average density
of $\rho_{{\mathrm{c}}}$ that is less mobile.
The six values of $p$ correspond
to three possible hexagonal backbones of a triangular lattice
in conjunction with two choices for the sign of 
$\rho_{{\mathrm{hx}}} -\rho_{{\mathrm{c}}}$.
Note that this spatial order is accompanied
by a very slight deviation of the total density $\rho$ from $1/2$
(which survives in the $T=0$ thermodynamic limit), with the
sign of deviation given by that of $\rho_{{\mathrm{hx}}} -
\rho_{{\mathrm{c}}}$.
We have also monitored these histograms at larger $U\lesssim40$. We find
that $\theta_K$ remains strongly pinned to $-2\theta_n$, and
while the pinning of $\theta_n$ and $\theta_K$ individually
does become weaker, the basic picture of the supersolid
state remains the same.

{\it{Landau theory:}}
Much of this picture of the supersolid phase can be understood within
the framework of a Landau theory written in terms of the order parameters
$\psi_n$ and $\psi_K$ (while it is not necessary to do so,
we find it convenient to explicitly include
$\psi_K$ in our description).
For our purposes here, it suffices to consider
only the `potential energy' terms of the Landau theory and leave out
all fluctuation terms that involve spatial and time derivatives, or
couplings to the superfluid order parameter, although these can also be
straightforwardly written down. Terms in the Landau
theory are constrained by the requirement of invariance under all the
symmetries of the system. The action of these on our order parameters
is simple to state: Under both lattice translations $T_0$ and $T_2$
we have $\psi_{n} \rightarrow e^{2\pi i/3} \, \psi_n$, $\psi_{K} \rightarrow e^{2\pi i/3} \, \psi_K$, while under a $\pi/3$ rotation about a $A$ sublattice site, 
we have  $\psi_{n} \rightarrow \psi_{n}^{*}$,
$\psi_{K} \rightarrow \psi_{K}^{*}$. Finally, $\psi_n$ is odd under particle-hole
transformations, while $\psi_K$ is even. Terms consistent with
these symmetries at $\mu=0$ give, up to sixth order
\be
S_{pot}(\psi_n, \psi_K)  =  f(|\psi_n|^2,|\psi_K|^2) +
 c_{\theta_n} (\psi_n^6 + {\psi_n^{*}}^6) \nn \\
+ c_{\theta_K} (\psi_K^3 + {\psi_K^{*}}^3) 
+ c_{nK}(\psi_n^2 \psi_K + {\psi_n^*}^2 \psi_K^*) \, .
\label{landau}
\ee
As usual, spatial symmetry breaking corresponds to the function $S_{pot}$ developing
a minimum at a nonzero value of $|\psi_n|$. The detailed nature of the ordering
is determined by the signs of coefficients $c$ which fix the relative as
well as absolute phases of the two order parameters. The results
shown for $U = 10$ in Fig.~\ref{thetahistogram} can be modeled by taking all $c$ negative, and this translates to the schematic picture of the phase
shown in Fig.~\ref{phases}. In addition, the very slight
$\theta_n$ dependence of $\rho-1/2$ can be modeled\cite{Melkounpub}
by the presence of a coupling term 
$(\rho-1/2)(\psi_n^3+{\psi_n^*}^3)$ with a tiny positive coefficient.

{\it{Discussion:}}
We have thus established the presence of a persistent low-temperature supersolid phase on the triangular lattice. The remarkable
stability of this phase is in contrast to the relatively small window of
parameters within which supersolids have been previously seen on
the square lattice.\cite{Hebert0,Troyer1} Indeed, a smooth extrapolation
of our data indicates that the supersolid phase persists in the
$U/t \rightarrow \infty$ limit.
Our results thus throw up
an interesting possibility for the observation of this phase in
atom-trap experiments. It would therefore be useful to map out
the finite temperature diagram for $U > U_c$. In the absence of any
coupling between spatial and superfluid order parameters, superfluidity
would be lost by a Kosterlitz Thouless (KT) phase transition, while
crystalline order would be lost via two KT phase transitions with
an intermediate power-law ordered crystal phase\cite{Jose0} analogous to
that seen in the transverse field Ising antiferromagnet on the same
lattice.\cite{Isakov0,Blankshtein0}
The coupling between the two order parameters is expected to
modify the detailed nature of the finite temperature phase diagram;
this will be reported on separately.\cite{unpublished}
Our picture of the supersolid state indicates that
it is not destabilized by doping, and this is currently
under investigation.\cite{unpublished}
Finally, it would also be interesting to study the stability of the supersolid
upon relaxation of the hard-core constraint.

{\it{Acknowledgements:}} One of us (KD) would like to acknowledge stimulating
discussions with A.~Paramekanti and A.~Vishwanath that led us to the work
described here,
and thank M.~Barma, S.~Gupta and T.~Senthil for useful suggestions.
We are also grateful to D.~Dhar for very insightful suggestions, and M.~Barma
and D.~Dhar for a critical reading of the manuscript.
Computational resources of TIFR,
and a student scholarship (DH) from the TIFR Alumni Fund are gratefully
acknowledged. During completion of
this work we became aware of parallel work\cite{Melkounpub} with
similar conclusions, and would like to thank the authors, especially
L.~Balents, for correspondence comparing and contrasting our results.

\end{document}